\documentclass[%
 reprint,
 amsmath,
 amssymb,
 aip,
 apl,
 noeprint,
 floatfix
]{revtex4-2}

\usepackage[utf8]{inputenc}
\usepackage[T1]{fontenc}
\usepackage{graphicx}
\usepackage{subfigure}
\usepackage{xcolor}
\usepackage{amsmath}
\usepackage{amssymb}
\usepackage{siunitx}
\usepackage{import}
\usepackage{newtxtext,newtxmath}
\usepackage{microtype}
\usepackage{hyperref}
\hypersetup{colorlinks=true, allcolors=blue}

\newcommand{\rhol}{\rho^{\ell}}
\newcommand{\rhon}{\rho^{\rm n}}
\newcommand{\rholi}[1]{\rhol_{\bs #1}}
\newcommand{\rhoni}[1]{\rhon_{\bs #1}}
\newcommand{\Ml}{M^{\ell}}
\newcommand{\Mn}{M^{\rm n}}

\newcommand{\epsll}{\epsilon_{\ell\ell}}

\newcommand{\epswl}{\epsilon_{{\rm w} \ell}}

\newcommand{\Mli}[1]{\Ml_{\bs{#1}}}
\newcommand{\Mni}[1]{\Mn_{\bs{#1}}}

\newcommand{\grad}{\bs \nabla}
\renewcommand{\div}{{\bs \nabla} \cdot}

\newcommand{\fd}[2]{\frac{\delta{#1}}{\delta {#2}}}
\newcommand{\bs}[1]{{\mathbf #1}}
\newcommand{\cF}{\mathcal{F}}

\begin{document}

\title{Changing the flow profile and resulting drying pattern of dispersion droplets via contact angle modification}

\author{Carmen Morcillo Perez}
\affiliation{SUPA School of Physics and Astronomy, The University of Edinburgh, Edinburgh, EH9 3FD, Scotland, United Kingdom}
\author{Benjamin D. Goddard}
\affiliation{School of Mathematics and Maxwell Institute for Mathematical Sciences, The University of Edinburgh, Edinburgh, EH9 3FD, Scotland, United Kingdom}
\author{Theo Reid}
\affiliation{SUPA School of Physics and Astronomy, The University of Edinburgh, Edinburgh, EH9 3FD, Scotland, United Kingdom}
\author{Schon Fusco}
\affiliation{Universidad de Granada, Department of Applied Physics, Granada, 18071, Spain}
\author{Andrew B. Schofield}
\affiliation{SUPA School of Physics and Astronomy, The University of Edinburgh, Edinburgh, EH9 3FD, Scotland, United Kingdom}
\author{Miguel A. Rodr\'{i}guez-Valverde}
\affiliation{Universidad de Granada, Department of Applied Physics, Granada, 18071, Spain}
\author{Marcel Rey}
\affiliation{SUPA School of Physics and Astronomy, The University of Edinburgh, Edinburgh, EH9 3FD, Scotland, United Kingdom}
\affiliation{University of M\"{u}nster, Institute of Physical Chemistry, Corrensstr. 28/30, 48149 M\"{u}nster, Germany}
\author{Job H. J. Thijssen}
\email[]{j.h.j.thijssen@ed.ac.uk}
\affiliation{SUPA School of Physics and Astronomy, The University of Edinburgh, Edinburgh, EH9 3FD, Scotland, United Kingdom}

\begin{abstract}
Spilling tea or coffee leads to a tell-tale circular stain after the droplet dries, known as the “coffee ring effect”. The evaporation of suspension droplets is a complex physical process, and predicting and controlling the particle deposit patterns from sessile droplet evaporation are essential for many industrial processes, such as ink-jet printing and crop-care applications. In this article, we systematically investigate the effect of surface wettability on the evaporation dynamics of a particle-laden droplet, focussing on the contact line stick-slip behaviour, the hydrodynamic flow of the suspended particles, and the resulting particle deposit after evaporation.
We use substrates with different wettabilities, ranging from hydrophilic to hydrophobic, and quantify the internal flow during the evaporation by tracking (fluorescent) tracer particles.
We find that the internal flow shifts from a predominantly outward flow towards the contact line for low contact angles to an inward flow for large contact angles. 
Correspondingly, the particle deposit transitions from the typical coffee-ring pattern to a central stain as the substrates's hydrophobicity increases. 
Finally, we corroborate these experimental findings with dynamic density functional theory, modelling the droplet evaporation process and stick-slip behaviour of the contact line. Our investigation suggests that the wettability of the substrate can significantly alter hydrodynamic flow within drying droplets, thereby affecting the resulting particle deposit.
\end{abstract}

\maketitle

\section{Introduction}

    When a drop containing disperse solids dries it leaves a solid deposit, often taking the characteristic form of a ring-like pattern, known as the coffee-ring effect (CRE).\cite{Deegan1997,Hu2002,Mampallil2018} Robert Deegan \textit{et al.} were the first to attribute this phenomenon to the pinning of the contact line of the droplet to the substrate. \cite{Deegan1997} They showed that liquid evaporates faster from the edges of the drop compared to its apex, inducing an outward capillary flow that carries the dispersed material to the edge of the droplet and therefore leads to a ring-like deposit. \cite{Deegan1997, Deegan2000-1, Deegan2000-2} 
    
    Suspension droplet evaporation is relevant for a variety of applications such as the fabrication of transparent conductive electrodes,\cite{Layani2009} nanochromatography,\cite{Wong2011} ink-jet printing, \cite{Son2019, Talbot2014} forensic investigations, \cite{Smith2018} crop-care applications \cite{Hunsche2012, Yu2009, Li2020} or diagnostics.\cite{Gulka2014,devineau2016protein} However, most applications typically require a specific dried pattern, such as a uniform deposit, hence a wide range of strategies have been explored to overcome the CRE. For example, the addition of soluble surfactants \cite{Erbil2015, Still2012,Marin2016} or salts, \cite{Marin2019, Pradhan2016, Nguyen2013} changing the particle size, \cite{Parthasarathy2021} or using various liquids \cite{Pyeon2021, Hu2006} can manipulate the hydrodynamic flow within the droplet and at least partially counteract the capillary flow and therefore improve the uniformity of the deposit.   
    The capillary flow can also be reduced by an increase in viscosity due to the addition of soluble polymers\cite{Cui2012} or by a gelation of the liquid. \cite{Talbot2014,Li2018}
    Furthermore, in recent work involving some of the authors, it was demonstrated that the CRE can be overcome by coating the suspended particles with surface-active polymers.\cite{Rey2022}
    More complex deposition patterns were found by manipulating the temperature, \cite{Ta2016-2, Parsa2015, Y.Li2015} electrowetting \cite{Eral2011} or confinement.\cite{Mondal2020}

    Important aspects governing the drying behaviour of sessile droplets are the substrate's properties, such as its roughness \cite{Kabi2021, S.Li2015, Dicuangco2014, Brunet2012, Lafuma2011} and wettability.\cite{Iqbal2018, Lin2015, Y.Li2013, Kumar2021}
    As a particular example, the wettability of different plant species varies from hydrophilic up to superhydrophobic.\cite{Fang2019, Papierowska2018, Fernandez2017} 
    For crop-care applications, it is therefore relevant to understand how the drying behaviour is affected by the different wettability of the specific plant leaves.\cite{Hunsche2012,Yu2009} 
    For (super)hydrophobic substrates, suspension droplets typically dry more into a point-like shape formed in the centre of the droplet instead of a coffee ring,\cite{Iqbal2018, Lin2015,han2024} even if the droplet is pendant,\cite{Kumar2021} though contact-angle hysteresis should also be considered.\cite{Y.Li2013} For example, a change in the contact angle hysteresis was observed for (super)hydrophobic substrates,\cite{Lin2015,Yang2020,han2024} with a constant contact angle mode instead of a constant contact line mode, which was attributed to the reduced pinning of the contact line.

    Gaining further insight into the CRE from modelling of particle-laden droplets is challenging due to the interplay between thermodynamic effects, such as phase transitions and hydrodynamic effects, including fluid flow within the droplet. \cite{ZCD15}  However, a key advantage is the ability to disentangle various physical properties such as surface wettability and slip.  
    Theoretically, a range of CAs (10, 90, 170, 180$^{\circ}$) were explored to study the Constant Contact Angle (CCA) and Constant Contact Line (CCL) extreme modes of evaporation, however no particles were present in their calculations. \cite{Stauber2015}
    For CA between 0 and 90$^{\circ}$, the Stokes flow has also been determined numerically. \cite{Hu2005}
    For large CAs they predict a positive Marangoni number that leads to an inward radial flow along the droplet surface. This flow is predicted to be outward at small CAs.
    The effect of the initial CA on the evolution of the flow pattern inside an evaporating sessile drop has also been numerically investigated. \cite{Chen2020} In this case, both Marangoni flow and natural convection were found to be key for large CAs.
    
    Here we present a systematic study of the drying behaviour of colloidal suspensions as a function of substrate wettability. 
    We use (confocal) fluorescence microscopy combined with the tracking of tracer particles to quantitatively characterise the hydrodynamic flow within the droplets during the evaporation process, as well as laser-sheet imaging to visualize the flow. 
    Interestingly, while we observe a primarily outwards flow along the substrate for low contact angles, as expected from previous literature, \cite{Deegan1997, Parsa2018, Zang2019} the observed flow for higher contact angles becomes predominantly inwards. 
    Additionally, upon increasing the substrate hydrophobicity, the dried deposit gradually changes from the typical coffee-ring to a dot-like deposit.
    This reveals that modifying the substrate wettability could be a new way of obtaining a desired deposit without the need to alter the liquid formulation or ambient conditions.
    Finally, we complement our experimental results with dynamic density functional theory simulations to verify the droplet evaporation behaviour and to elucidate the role of the stick-slip behaviour of the contact line.

\section{Materials and methods}

    \subsection{Dispersion preparation}
        Unless stated otherwise, droplets used were 1 $\mu$L in volume and contained 0.1 \%w of red fluorescent carboxylate-modified polystyrene latex beads (mean particle diameter 1.90 $\mu$m, Sigma-Aldrich) and 0.0005 \%w of yellow-green fluorescent carboxylate-modified polystyrene latex beads (mean particle diameter 1.90 $\mu$m, Sigma-Aldrich) in distilled and de-ionized water (18 M$\Omega$cm, Milli-Q RG Opak).
        The dispersion was centrifuged and the surface tension of the supernatant water was measured to be 73 mN m$^{-1}$  (with a Kr$\mathrm{\ddot{u}}$ss DSA100 drop tensiometer model 65 FM40Mk2) which aligns well with literature values for a clean air-water interface at 21 $^o$C, \cite{Vargaftik1983} suggesting that there was no interfacially active contamination in the supernatant.
        The beads were imaged with a confocal microscope (Zeiss LSM 700) and Scanning Electron Microscope (SEM, JEOL, 6010 LV, 20kV, 15 nm gold-coated samples) to assess overall round appearance, and to confirm that particle size aligns with supplier information.

        For confocal-microscopy measurements, fluorescent and non-fluorescent St\"{o}ber silica particles were synthesized in-house.\cite{Imhof1996,Blaaderen1992} The average diameter of the unlabelled (UL) particles was 0.386 $\mu \mathrm{m}$ with a polydispersity of 6.4\%, as determined by scanning electron microscopy (SEM, JEOL, 6010 LV, 10 kV, see ESI); static light scattering (SLS, ALV, LSE-5004) gave an average diameter of 0.438 $\mu \mathrm{m}$ with a polydispersity of 6.3\%. The fluorescently labelled (FL) particles were labelled with fluorescein isothiocyanate (FITC), with 3-aminopropyltriethoxysilane (3-APS) as the linking molecule. The FL particles had an average diameter of 0.603 $\mu \mathrm{m}$ and a polydispersity of 4.3\% (SEM, Au-coated, see ESI); the average diameter according to light scattering (Beckman Coulter, LS 13 320) was 0.643 $\mu \mathrm{m}$. The difference between the average diameter from SEM (in vacuum) and light scattering (in solution) is in line with expected shrinkage of St\"{o}ber silica particles in the vacuum of the SEM.\cite{Imhof1999} Before use, silica particles were washed by repeated centrifugation and redispersion in distilled water, until the supernatant was colourless and had a pH of 7. For sample preparation, the particles were diluted with distilled water to an overall concentration of 0.0002 wt-\% labelled particles and 0.4 wt-\% unlabelled particles.

    \subsection{Substrate preparation}
        
        For the polystyrene dispersions, four types of glass substrates were prepared in order to study different water contact angles.
        All glass substrates (Cat. No. 7101, Scientific Glass Laboratories ltd., Tunstall) were cleaned prior to use or functionalisation.
        
        For the cleaning stage the slides were first submerged in a 5 \% Decon 90 (R) cleaning solution (used as sold, Decon Laboratories Limited) in distilled and deionised water (18 M$\Omega$cm, Milli-Q RG Opak) and sonicated for 10 minutes in an ultrasonic cleaner (USC200T, VWR ultrasonic cleaner, frequency 45 kHz, max. output power 120 W, Ultraschall Sonic Ultrasons). These were then rinsed with distilled and deionised water. Finally, the glass slides were plasma-treated (Zepto, model 2, Diener Electronic) for 10 seconds at 35 W.
        
        For the silanisation step, the cleaned substrates were submerged in a solution of 40 mL ethanol ($\geq$ 99.8 \%, Sigma-Aldrich) and 4 mL aqueous ammonia solution (35 \%, Fisher Chemicals).
        One slide was functionalised with 50 $\mu$L of 1H,1H,2H,2H-Perfluoro-octyltriethoxysilane (FOTS) (98 \%, Sigma-Aldrich) giving initial CAs of $105.3 \pm 1.9^{\circ}$.
        Two slides were functionalised with 1 mL and 3 mL of 1,1,1,3,3,3-Hexamethyldisilazane (HMDS) (99 \%, Fluka Analytical), giving initial CAs of $54.6\pm 1.3^{\circ}$ and $95.2\pm 2.6^{\circ}$, respectively.
        The glass slides were submerged in the mixture for 72 hours with continuous stirring. The samples were then cleaned with ethanol and sonicated to remove potential silane multilayers.
        The final slide was kept pristine after the cleaning stage, giving initial CAs of $37.7 \pm 1.0^{\circ}$.

        For confocal-microscopy measurements, three different substrates were used: glass, plastic (polystyrene) and silanised glass. Glass slides (Sail Brand, 1 to 1.2 mm thick) were cleaned with ethanol (Sigma-Aldrich, $\ge 99.8$\%) and a paper towel. Plastic Petri dishes (Thermo Scientific, Sterilin, standard 90 mm) were also cleaned with ethanol (Sigma-Aldrich, $\ge 99.8$\%) and a paper towel. To obtain silanised glass, a glass slide was cleaned with ethanol and subsequently submerged for 10 minutes in silanisation solution 1 (Sigma-Aldrich, $\sim 5$\% dimethyldichlorosilane in heptane); the slide was then cleaned with distilled water. The initial contact angle of a water droplet on the substrates was measured using a tensiometer (see below): $\mathrm{CA_i} = 32.0 \pm 0.5^{\circ}$ for glass, $\mathrm{CA_i} = 46.6 \pm 0.5^{\circ}$ for plastic, and $\mathrm{CA_i} = 101.6 \pm 1.9^{\circ}$ for silanised glass.
        
    \subsection{Side-view imaging}
        A 1 $\mu$L drop of the polystyrene dispersion was deposited on the substrate using a glass syringe (Hamilton, 450 $\mu$L) with a stainless steel needle (luer-lock, 0.499 mm diameter) connected to a software-controlled dosing system.
        To capture image sequences of the drop we used a Drop Shape Analyser (DSA100, Kr$\mathrm{\ddot{u}}$ss, model 65 FM40Mk2) with a charge-coupled device camera (Stingray) controlled by a PC. The focal plane was positioned across the middle of the droplet when generated. The image sequences were analysed using an in-house Matlab code using a fitting method based on a circle fit by Taubin. \cite{Taubin1991}
        All the substrate types were measured at least three times, by depositing a new droplet in a different spot on the substrate. The mean of the CA was used for the analysis and plotting. 

        For the confocal-microscopy system, droplets of 2 $\mu \mathrm{L}$ volume were deposited on the substrate using a pipette. The Drop Shape Analyser was used to measure the initial contact angle for each substrate type using a Young-Laplace fit; cited $\mathrm{CA}_i$ values are the average over three separate droplets. The same instrument was used to measure contact angles and base diameters of droplets over the duration of evaporation.
                   
    \subsection{Fluorescence microscopy}
        A 1 $\mu$L drop was carefully deposited onto the substrate with a micropipette (0.5 - 10 $\mu L$, TopPette Pipettor). A semi-automated upright epi-fluorescence microscope (Nikon Eclipse E800) with a camera (QImaging Retiga 2000R Fast 1394 Cooled) and a Nikon CFI Plan Fluor objective (10$\times$ magnification, NA 0.30) was used to take images of the drop every 2 seconds throughout the evaporation process, with the focal plane just above the solid-liquid interface.
        The illumination filter sets used were EGFP excitation filter (Chroma ET470/40x) and EGFP emission filter (Chroma ET525/50m) for the particle tracking; and mCherry excitation filter (Chroma ET572/35x) and mCherry emission filter (Chroma ET632/60m) for the imaging of the dried deposit.
        The system was enclosed in an incubator chamber at all times for monitored temperature (19 $\pm$ 2$^{\circ}$) and humidity (46 $\pm$ 2 \%).

    \subsection{Confocal microscopy}
        Droplets of 2 $\mu \mathrm{L}$ volume were deposited on the substrate using a pipette. They were imaged during evaporation in a time sequence of 3D scans using a confocal microscope (Leica Microsystems, DMi8 TCS SP8) with a dry objective ($10\times$, 0.3 NA). Time series were taken near the edge of the droplet at three different stage heights, $h$, within the same evaporating droplet: close to the substrate at $h = 19 \ \mu \mathrm{m}$, in the middle at $h = 113 \ \mu \mathrm{m}$ and close to its apex at $h = 188 \ \mu \mathrm{m}$ (see ESI); by stage height we mean that we have not corrected the height above the substrate for axial distortion due to refractive index mismatch.\cite{Diel2020} FITC was excited using a 488 nm laser and the emitted light was captured with a photomultiplier detector. The resonant scanner was used at 8,000 Hz, resulting in a 2.17 second time interval between images at the same height above the substrate.

        To obtain micrographs of final deposits for the confocal-microscopy system, similar droplets were deposited on similar substrates and imaged in brightfield mode with an upright microscope (Olympus, BX50, $5\times$ objective) with a mounted camera (QImaging, QICAM). For the untreated glass substrate, the deposit was captured in panels, which were subsequently stitched together using the Fiji (ImageJ) Stitching plugin.\cite{Preibisch2009} A background correction (rolling ball) was applied to all (composite) deposit micrographs.

    \subsection{Laser-sheet imaging}
        To visualize the cross-sectional flow inside the sessile droplets, a very low concentration (0.0005 w/v\%) of fluorescent tracers (1.14 $\mu$m-polystyrene spheres, Microparticles GmbH) was added to 3 $\mu$L-water droplets containing 0.05 w/v\% of 1.02 $\mu$m-PMMA spheres (Microparticles GmbH). Droplets were placed on each substrate by using a handheld pipette. A laser sheet technique, based on a 532 nm-CW laser source (SDL-532-1500T, Shanghai DreamLaser Technology), was employed for flow field visualization. The laser beam was collimated with two spherical lenses and expanded in one direction with a cylindrical lens to form a 0.3 mm thick sheet, then illuminating the droplet central plane. Droplets images were captured with a high-speed camera (Phantom Miro C110) equipped with a variable magnification objective (MLM3X-MP, Computar). A dichroic filter (FD1R, Thorlabs) was used to block background and scattered light, allowing only the fluorescent light from the tracer particles to reach the camera. An Arduino board acted as a modulator, converting the continuous laser into a pulsed beam. Particle Image Velocimetry (PIV) was performed on the captured fluorescent signals using PIVlab.\cite{Thielicke2014,Thielicke2021} Frames were taken in sets of four at 15-second intervals, followed by a 60-second pause before repeating the sequence. This protocol provided detailed insights into the droplet internal flow dynamics over time.

    \subsection{Image analysis}\label{sec:imganalysis}

        \subsubsection{Fluorescence data}

            The fluorescence microscopy image sequences were analysed in Python with a particle tracking code based on the TrackPy module (v0.6.1).\cite{Allan2019}
            Tracking parameters were optimized manually by trial and error, and checked visually for the most acceptable identification of the particles. The diameter was set to 15 pixels and the minimum separation between features to 5 pixels. The maximum displacement for a particle to move between frames was kept to 15 pixels and the memory to 3 frames, which is the number of frames to keep the particles' ID if not located.
            
            Tracked particles were filtered by appearance: the minimum integrated brightness for the features was set to 200 and the maximum radius of gyration of brightness to 8 pixels; the maximum eccentricity was chosen to be 0.5. A threshold of 2.5 was chosen in order to clip band-pass results. The "noise size", or width of Gaussian blurring kernel, was left as the default of 1 pixel. Finally, the data has been filtered to show only trajectories that last for at least two frames. The same parameters were used to analyse all four sequences shown.
            
            Assuming that the particle tracking resolution is 1/10 of a pixel at best, \cite{Crocker1996} particles with an apparent velocity of 0.037 $\mu$m/s or smaller should be considered stuck.
            Tracking particles outside of the droplet edge, which should be stuck, we get an average velocity of 0.034 $\mu$m/s, which is in good agreement with the resolution argument.
            As such, displacements less than 0.074 $\mu$m per 2 s frame interval are ignored.
                
        \subsubsection{Confocal data}

            To analyze the confocal-microscopy data, we used  the ``TrackMate'' plugin \cite{Ershov2022} in Fiji. \cite{Schindelin2012} The Laplacian of Gaussian (LoG) detector was found to be optimal for detecting the silica colloidal particles in this experiment with reasonable computation time. The estimated object diameter was set to 10 pixels and the quality threshold was set to 0. After object detection, initial thresholding was set to 0. The filter settings, ‘Std intensity ch1’ and ‘Contrast ch1’ were set to 5.14 and 0.11 respectively. This was optimized by observation, but not every particle in each frame was detected. However, the large number of particles detected should represent the general movement of particles providing an understanding of the flow within the evaporating droplets. For tracking, the Linear Assignment Problem \cite{Jaqaman2008} (LAP) tracker was selected. The default tracking parameters were used and frame-to-frame linking was set to 15 pixels.
            
            The exported particle tracks were analyzed in Python using the Pandas library.\cite{mckinney2010pandas} Particles that had been recorded for a single frame only were discarded. The radial distance travelled is defined as positive when a particle moves outward toward the contact line, and negative when moving towards the centre of the droplet. The centres of the droplets were determined visually.
            
            All recorded distances less than 0.08 $\mu$m between consecutive frames corresponding to a single particle were removed since these particles were considered stationary. For each frame, the distance travelled by every particle was summed to give a total distance travelled in the frame, which was then normalised by the number of particles tracked in the frame.

    \subsection{DDFT model}

        DDFT describes the time evolution of ensemble-averaged densities for the liquid ($\rhol$) and particles ($\rhon$) in a droplet.  These densities evolve according to a pair of coupled partial differential equations (PDEs):
        \begin{align}
        	\partial_t \rhol(\bs{x},t) &= \div \left[ \Ml(\bs{x},\rhol,\rhon) \rhol \grad \fd{\cF[\rhol,\rhon]}{\rhol} \right] \\
        	\partial_t \rhon(\bs{x},t) &= \div \left[ \Mn(\bs{x},\rhol,\rhon) \rhon \grad \fd{\cF[\rhol,\rhon]}{\rhon} \right].
        \end{align}
        Here $\bs{x}$ is the spatial coordinate; $t$ is time; $\Ml$ and $\Mn$ are the mobility coefficients for the liquid and nanoparticles, respectively; and $\cF$ is the Helmholtz free energy. 
        Note that typically $\rhon \ll \rhol$ in the initial condition, representing a dilute mixture.  The boundary conditions are no-flux at the substrate, evaporative at the top of the domain, and periodic on the boundaries perpendicular to the substrate.  Our initial conditions are smoothed spherical caps. See ESI for further details of the model. 
        
        Our model is taken from previous work on the evaporation of particle-laden droplets. \cite{CSA17} In Chalmers \emph{et al}.,\cite{CSA17} the dynamics are discretised onto a lattice, with the value of $\rhol$ at a lattice site $\bs{i}$ denoted $\rholi{i}$, and analogously for other quantities, giving
        \begin{equation}
        	\partial_t \rholi{i} = \div \left[ \Mli{i} \rhol \grad \fd{\cF}{\rholi{i}} \right], \quad
        	\partial_t \rhoni{i} = \div \left[\Mni{i} \rhon \grad \fd{\cF}{\rhoni{i}}  \right].
        \label{DDFT}
        \end{equation}
        The Helmholtz free energy is approximated using a mean field approach. See references within Chalmers \emph{et al}.\cite{CSA17} for a justification of this, and the ESI for the precise form of $\cF$.  We take the same approach as in Chalmers \emph{et al}.\cite{CSA17} and average the equations into two dimensions with corresponding coefficients given in the ESI.  Note that the free energy introduces an `effective advection' term, where the dynamics of the nanoparticles are driven by the water density (and vice-versa).
        
        It remains to describe the mobilities in equation \eqref{DDFT}, as well as the boundary conditions for the lattice dynamics.
        For the mobilities, we take
        \begin{equation} M_\bs{i}^c = \begin{cases}
        m_\bs{i}^c
        \begin{pmatrix}
        s & 0 & 0 \\
        0 & \nu & 0 \\
        0 & 0 & s
        \end{pmatrix}
        & j = 1 \\
        m_\bs{i}^c
        \begin{pmatrix}
        1 & 0 & 0 \\
        0 & 1 & 0 \\
        0 & 0 & 1
        \end{pmatrix}
        & \rm{otherwise} \\
        \end{cases}
        \end{equation}
        Here $s$ and $\nu$ determine the slip at the substrate, with $s$ controlling motion parallel to the substrate ($s=0$ for a no-slip substrate), while $\nu$ prevents `hopping' of particles; typically $\nu \ll 1$. If $\nu \approx 1$ then the choice of $s=0$ no longer enforces a no-slip condition at the substrate as the liquid or particle migrates to the $j=2$ layer, moves parallel to the substrate, and then returns to the $j=1$ layer next to the substrate.  We assume that the liquid mobility is constant $m_\bs{i}^\ell = m_\ell$, while that of the particles depends on the density (i.e., the number of particles in a cell in the microscopic model) of the liquid: $m_\bs{i}^n = \tfrac{m_n}{2} \big[ \tanh(8 \rho_\bs{i}^\ell - 4) + 1 \big]$. This enforces that the particles do not move on a dry surface, but move freely when they are surrounded by liquid. The particular choice of $m_\bs{i}^n$ is arbitrary, but once again we have followed  Chalmers \emph{et al}.,\cite{CSA17} and the function smoothly interpolates between no particle motion when dry to free motion when `completely wet'.
    
        We note that, for a fixed initial condition, the size of the domain affects the speed of evaporation, with droplets in smaller boxes evaporating more rapidly. We choose the domain to be sufficiently large ($70 \times 140$) so that the only change in the dynamics observed by increasing or decreasing the size of the box is in the evaporation timescale -- the dynamics are unchanged after a time rescaling. We use the numerical method described in the Appendix of Chalmers \emph{et al}.\cite{CSA17} with $k_BT = 1$, $\epsll = 1.5$, and a timestep of $10^{-2}$ for no-slip dynamics ($s=0$) and $5\times 10^{-3}$ when $s=1$. For the two smallest contact angles, we deviate very slightly from the experimental means, due to the initial transient dynamics observed in the experiments. In particular, for contact angles [105$^\circ$, 95$^\circ$, 55$^\circ$, and 38$^\circ$], we choose $\epswl = [0.32,0.39,0.74,0.84]\epsll$, which corresponds to initial simulation contact angles of [105$^\circ$, 95$^\circ$, 52$^\circ$, 35$^\circ$]. Note that this choice has very little effect on the overall dynamics of the systems.

\section{Results and discussion}

\begin{figure}
        \centering
            \includegraphics[width=0.99\linewidth]{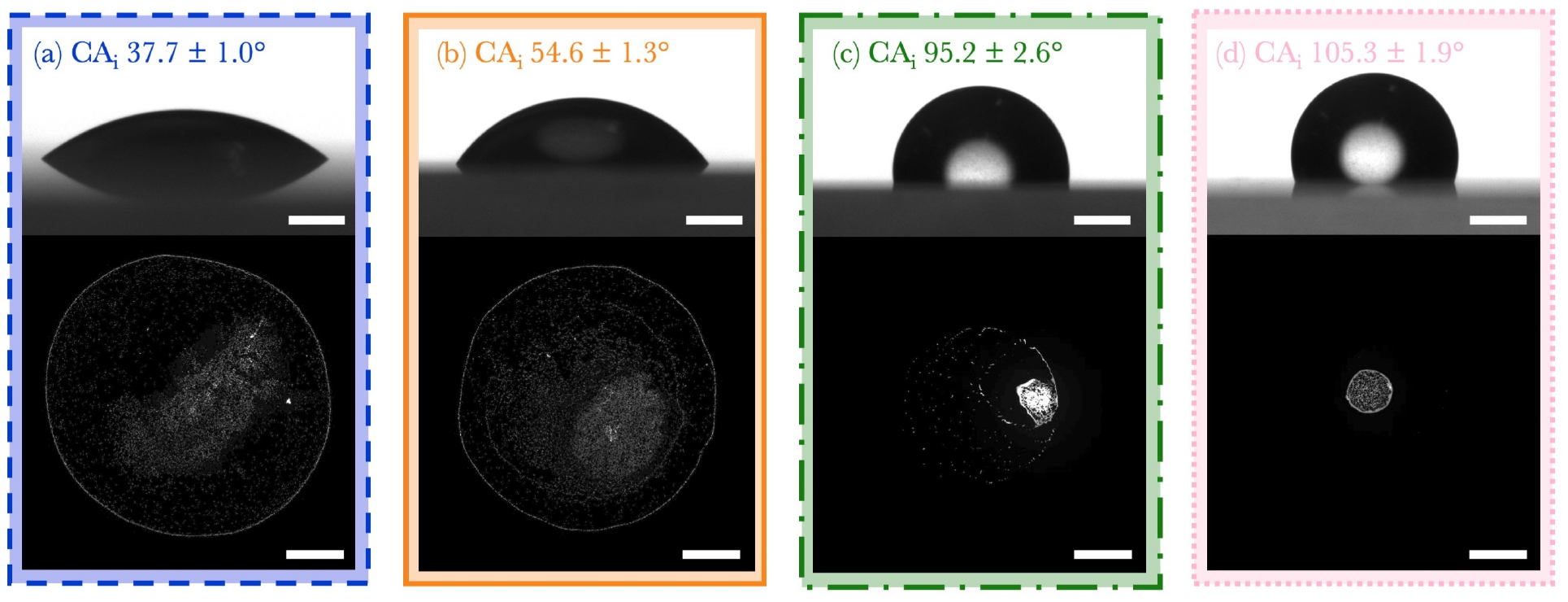}
        \caption{Droplet shape (top) and dried deposit (bottom) as a function of substrate wettability. Top: Side-view images of 1 $\mu$L water droplets containing fluorescent polystyrene particles on four different silanised glass substrates. Bottom: Stitched fluorescent microscopy images of the dried particle dispersion. $CA_i$ refers to the initial CA calculated using image analysis from a side-view image of the droplet. Scale bars: 0.5 mm.}
        \label{fig:sidetop}
    \end{figure}

    We first investigated the effect of the substrate's wettability on the droplet CA and the resulting drying pattern of polystyrene in water dispersions (figure \ref{fig:sidetop}). Although there is some variation between the repeats, the overall appearance of the deposits is qualitatively similar.
    
    A CRE is clearly observed for the substrate with smallest CA, $CA_i = 38^{\circ}$. The deposit for the substrate with $CA_i = 55^{\circ}$ also shows CRE; however, it is less clear as depinning is likely to have occurred leaving a second, less clear, smaller ring circumscribed. The dried pattern in the substrate with $CA_i = 95^{\circ}$ shows signs of multiple depinning steps and a continuous ring is not visible, but instead shows numerous discontinuous ring deposits and a bright spot of concentrated particles. The substrate with the biggest CA, $CA_i = 105^{\circ}$, stands out because of the small bright spot that is the main visible feature of the deposit. Upon closer inspection of this sample it is possible to see radial particle deposits, likely caused by a continuous inwards slip motion of the droplet, that leads to a very thin discontinuous ring-like deposit.
        
    Similar behaviour has been reported analytically in the literature. \cite{Masoud2009-1, Masoud2009-2} They calculated that the final dried pattern depends on the flow within the drop; this flow is dependent on the evaporative flux, shape of the free surface, and behaviour of the contact line. Subject to these factors, the flow can be inwards, outwards or a mixture. 
    We therefore decided that a more exhaustive analysis is needed so we recorded the evaporation process from a side-view and compared it to the results from the DDFT model.
       
    \subsection{Side-view imaging}
\begin{figure}
            \centering
                \includegraphics[width=0.99\linewidth]{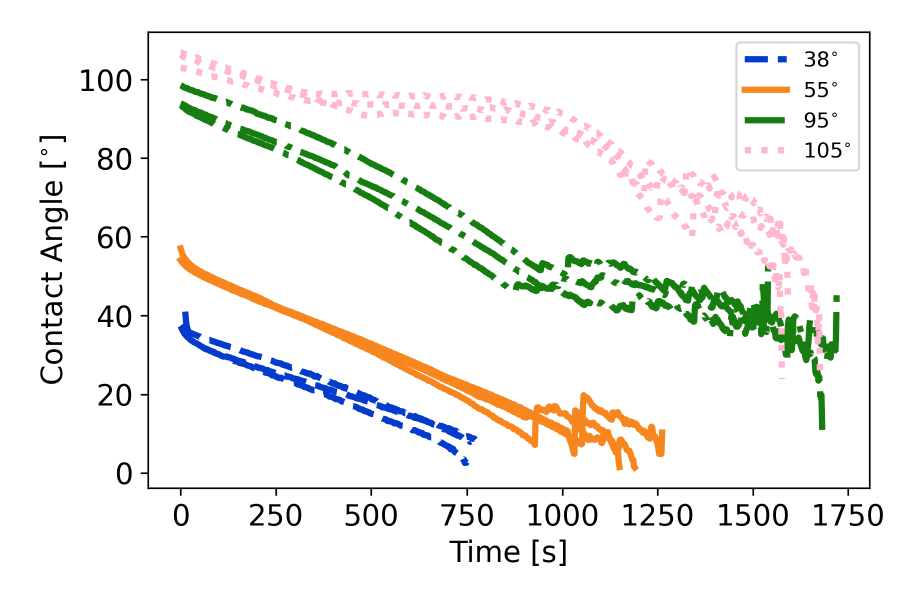}
            \caption{Graph showing the evolution of contact angle with evaporation time for water droplets with polystyrene particles on four different (silanised) glass substrates. The CA in the legend refers to the average initial CA at $t=0$.}
            \label{fig:CAvsTime}
        \end{figure}

        Image sequences of polystyrene-in-water dispersion droplets drying were recorded from the side for each of the four different substrates.
        The images at $time = 0$ can be seen in figure \ref{fig:sidetop} (top images) accompanied by the initial CA extracted.
        After taking three repeats per substrate we determined that although a small variation is observed, the overall shape of the evaporation curves are reproducible for the different substrates (figure \ref{fig:CAvsTime}).

        \begin{figure*}
            \centering
                \includegraphics[width=0.99\textwidth]{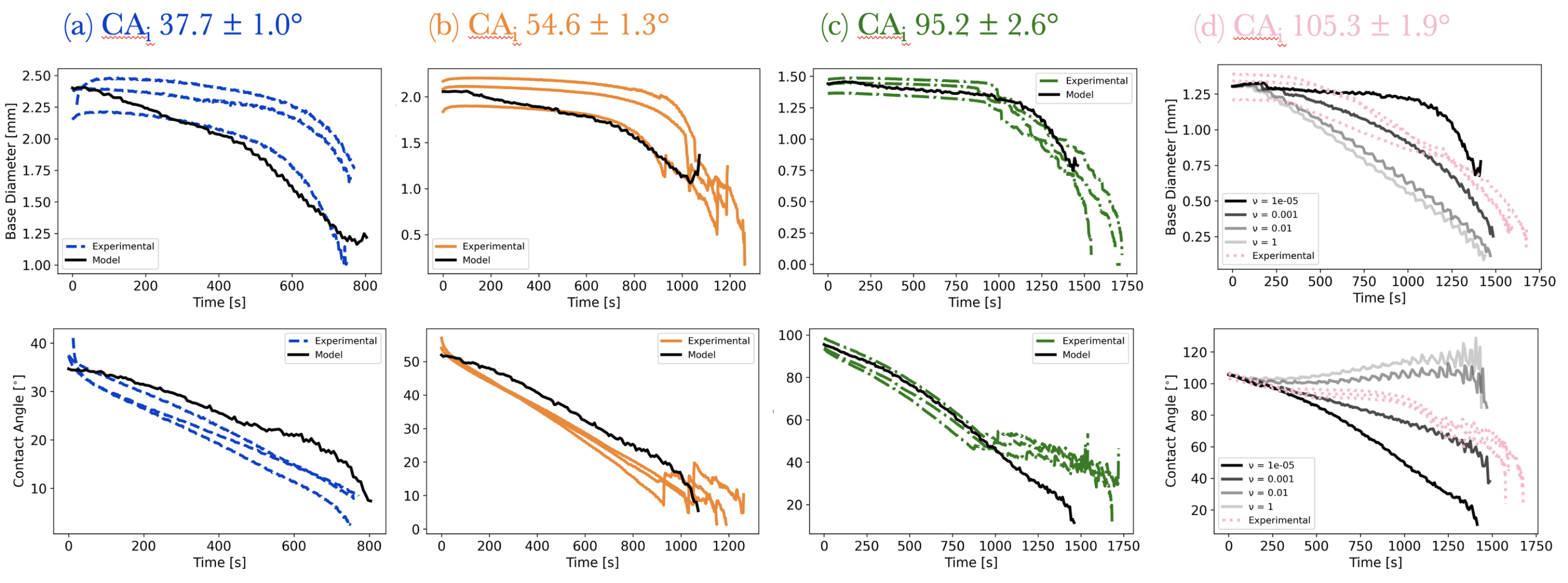}
            \caption{Graphs showing the evolution of base diameter (top) and contact angle (bottom) with evaporation time for four different (silanised) glass substrates. The CA in the legend refers to the average initial CA at $t=0$. The black lines represent the DDFT simulations for $\nu = 1 \cdot 10^{-5}$, unless otherwise specified.}
            \label{fig:CAvsTimeAll}
        \end{figure*}

        At first glance, substrates with $CA_i = 38^{\circ},~55^{\circ}$ and $95^{\circ}$ appear to have fairly similar wetting profiles (figure \ref{fig:CAvsTime}). For all three, the first stage of the evaporation seems to be of constant contact line (CL), as seen by the continuously decreasing CA.
        Similarly, the substrate with $CA_i = 105^{\circ}$ has an initial stage of constant CL, however this is shorter and followed by a constant CA stage, which is not seen for the other three substrates. The final part of evaporation appears as a mixed mode of both moving CL and changing CA for all substrates. These findings are in agreement with previous studies, where droplets also evaporate via a combination of “stick” and “slide” phases. \cite{Stauber2015,Lin2015,Yang2020} Occasionally, we observed asymmetric CL motion, where the droplet depinning only occurs on one part of the CL. 

        From this experiment we conclude that different initial CAs do not necessarily lead to a qualitatively different macroscopic contact-angle evolution (substrates with $CA_i = 38^{\circ},~55^{\circ}$ and $95^{\circ}$). Yet, because qualitatively different dried deposits have previously been observed (figure \ref{fig:sidetop}), analysis of the internal flows of the droplet will likely be key to understanding these differences (see section \ref{subsec:fluomicroscopy}--\ref{subsec:laserSheetImaging}).
        
        DDFT simulations have been performed to model the droplet evaporation process and the corresponding results are shown in figure \ref{fig:CAvsTimeAll}.
        Unless otherwise stated we use a `hopping' parameter  value of $\nu = 1 \cdot 10^{-5}$ (see above), giving good agreement with the experimental results where a constant contact line mode is dominant throughout the evaporation process (substrates with $CA_i =$ 38, 55 and 95$^{\circ}$). However, for the substrate with $CA_i = 105^{\circ}$, a range of $\nu$ parameters are presented, showing that this can be applied to tailor the stick-slip behaviour of the substrate. 
        In other words, the simulations can also reproduce the observed behaviour for $CA_i = 105^{\circ}$ by choosing an appropriate value for the parameter $\nu$. As a different silanising agent was used to prepare this substrate, it is perhaps not surprising that a different value of $\nu$ fits the experimental data better.  We note that $\nu$ likely depends on local surface properties, such as heterogeneities, and there is no mechanism in the model for switching $\nu$ over time, which would allow a more accurate reproduction of the experimental results for $CA_i = 105^{\circ}$; this is an interesting topic for future work.  Note that these simulations were performed with no particles, i.e., $\rho^n = 0$; simulations with low particle concentrations show essentially identical results except in the final stages of evaporation. Results including particles can be found in the corresponding research data.
        
        The model aims to enforce the constant CL condition by restricting the mobility of the water and particles in the cells closest to the substrate. This prevents movement parallel to the substrate and strongly limits movement perpendicular to the surface, preventing `hopping'.~\cite{CSA17} This seems to work well in the cases of large contact angles; see videos in corresponding data files. In contrast, in the case of small contact angles, the water forms a film in the layer above the substrate. This film layer is indeed fixed, but then the droplet evolves above this layer, leading to a more constant CA-like behaviour. This is likely a consequence of the increased water-substrate attraction versus the water-water interactions, compared to the larger contact angle cases; this is needed to ensure the correct equilibrium contact angle. 

    \subsection{Fluorescence microscopy}
    \label{subsec:fluomicroscopy}
        
        The droplet evaporation was recorded from the top using fluorescence microscopy to track tracer particles. An EGFP filter is used, to track only the green-dyed polystyrene beads (concentration = 0.0005 \%), although the total concentration of polystyrene particles is 0.1 \% (see methods for further information).  
        The image sequences are analysed in Python with a particle tracking code based on the TrackPy module. \cite{Allan2019} The particles are identified in each of the images, linked from frame to frame and filtered using an implementation of the Crocker-Grier linking algorithm, see section \ref{sec:imganalysis}. \cite{Crocker1996}  

        \begin{figure*}
          \centering
            \includegraphics[width=0.99\textwidth]{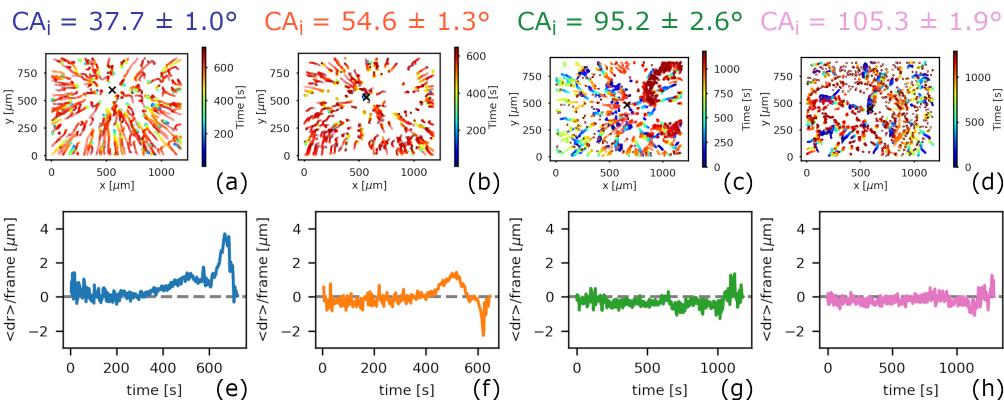}
            \caption{Top: Position of polystyrene tracer particles throughout droplet evaporation; the colour gradient indicates time and a black cross indicates the centre of the drop. The focal plane was located just above the substrate. Only tracks of at least 30 frames (60 s) are shown for clarity. Bottom: Average radial displacement of particles relative to the centre of the drop (positive/negative means movement outwards/inwards); displacements less than 0.074 $\mu$m were ignored (see text for details).}
            \label{fig:PTtrajectories}
        \end{figure*} 
        
        The location of tracked particles throughout the evaporation process is shown in figure \ref{fig:PTtrajectories} (top row), where the colour bar refers to the time elapsed. Most particles seem to have a radial trajectory, towards the centre or edge of the drop. In order to quantify this, the position of the centre of the drop is subtracted from the position of each particle and the average radial displacement per frame is plotted against time elapsed (figure \ref{fig:PTtrajectories} bottom row).
        It is important to note that the position of the droplet centre is chosen manually and kept the same for the whole evaporation process. If the CL depinning is asymmetric, the centre of the drop may change slightly over time. However, looking at figure \ref{fig:PTtrajectories} (top row), the chosen centre seems to be consistent for the particles' trajectories.
 
        For the pristine glass substrate ($CA_i = 38^{\circ}$), the predominant outward flow, especially near the end of the evaporation process (figure \ref{fig:PTtrajectories}(a)), agrees with observations made in the literature and suggests a coffee-ring formation due to capillary outward flow. \cite{Deegan1997, Parsa2018, Zang2019, Rey2022}
        For the $CA_i = 55^{\circ}$ substrate, this peak in radial displacement is markedly less pronounced (figure \ref{fig:PTtrajectories}(b)); the subsequent trough is caused by the drying front moving through the field of view.
        For the hydrophobic substrates (figure \ref{fig:PTtrajectories}(c-d)), the average radial displacement is almost constant over time, and on average slightly negative i.e.~towards the centre of the drop; the slight increase near the end of evaporation is again caused by the drying front (at which point the marked droplet centre is no longer appropriate). Note that these observations align with the tracer particle position plots: the tracks are more orange/red, longer and more symmetrical for low initial contact-angle substrates.

        It should be pointed out here that particle tracking is challenging in this system because particles are out of focus initially and imaging through the water droplets causes image artefacts, especially for droplets with larger contact angles. However, the overall trend that we observe is the following. For most of the samples, towards the beginning of the evaporation process, distorted particles are sometimes identified as multiple centres. In addition, for some of the samples, towards the end of the evaporation process, the movement of the CL looks like it is pushing some of the identified particles inwards. However, the inward flow is also seen at earlier stages of evaporation where the movement of the CL is unlikely causing this effect as depinning is not seen in side-view imaging (for example for the sample with $CA_i=$95$^{\circ}$). Finally, some particles are not tracked due to particle identification and filtering parameters. Hence, we have re-analyzed the data corresponding to Figure \ref{fig:PTtrajectories} using the TrackMate plugin in ImageJ and this confirmed the trends observed in Figure \ref{fig:PTtrajectories} (see ESI).

        To summarize, instead of an outward flow, typically associated with the coffee ring effect, we measure an inward flow during the earlier stages of drying for substrates with a high contact angle. Only towards the later stage of the drying, during the constant contact line mode, we measure an outward flow. This may explain the difference in drying behaviour observed in Fig. \ref{fig:sidetop}. 

\subsection{Confocal microscopy}
   
\begin{figure}
          \centering
            \includegraphics[width=0.99\linewidth]{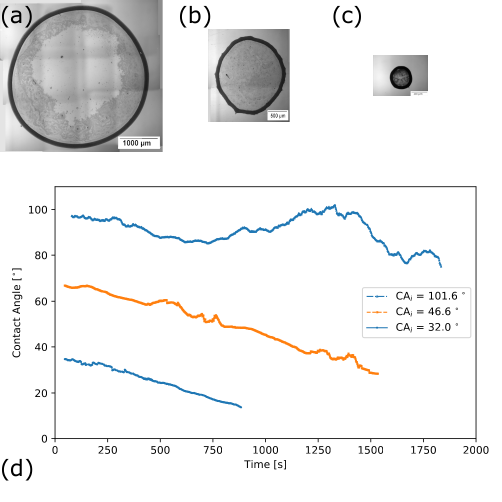}
            \caption{(a-c) Stitched bright-field microscopy images of dried deposits of silica particles on three different substrates: (a) non-functionalised glass ($\mathrm{CA}_i = 32.0^{\circ}$), (b) plastic ($\mathrm{CA}_i = 46.6^{\circ}$) and (c) silanised glass ($\mathrm{CA}_i = 101.6^{\circ}$). (d) Contact angle during evaporation of different but similar droplets on glass, plastic and silanised glass; a rolling average has been applied to each graph. Scale bars: (a) 1000 $\mu$m, (b) 500 $\mu$m and (c) 400 $\mu$m i.e.~panels (b) and (c) are to scale relative to panel (a).}
            \label{fig:confocalVelocityHeight}
        \end{figure}

Next, we corroborate the inward/outward flow pattern observed in fluorescence microscopy with both confocal microscopy and laser-sheet imaging. For the confocal-microscopy measurements, we used a silica particle dispersion and dried them on three different substrates with varying contact angles. First, we established that the final deposits and the side-view imaging for the droplets containing silica particles lead to similar results as previously presented for polystyrene particles.Figure \ref{fig:confocalVelocityHeight}(a-c) shows (stitched) bright-field micrographs for dried deposits of silica particles on three different substrates: (a) glass ($\mathrm{CA}_i = 32.0^{\circ}$), (b) plastic (polystyrene) ($\mathrm{CA}_i = 46.6^{\circ}$) and (c) silanised glass ($\mathrm{CA}_i = 101.6^{\circ}$). Qualitatively, the trend is similar i.e.~a clear CRE for low contact angle and a smaller spot for high contact angle, with a deposit in between those two for intermediate contact angle. Figure \ref{fig:confocalVelocityHeight}(d) shows the contact angle as a function of evaporation time for different but similar droplets on glass, plastic and silanised glass. Qualitatively, the behaviour is similar to above i.e.~continually decreasing CA for low and intermediate initial CA, and a significant constant CA stage for high initial contact angle. Notably, for high initial CA, the contact angle trend seems to remain constant throughout evaporation, whereas above there was a decreasing CA stage at early and late stage of evaporation; it could be the difference stems from the different interaction energies of silica and polystyrene with the substrates and/or the difference in silanisation procedure.

\begin{figure}
        \centering
            \includegraphics[width=0.99\linewidth]{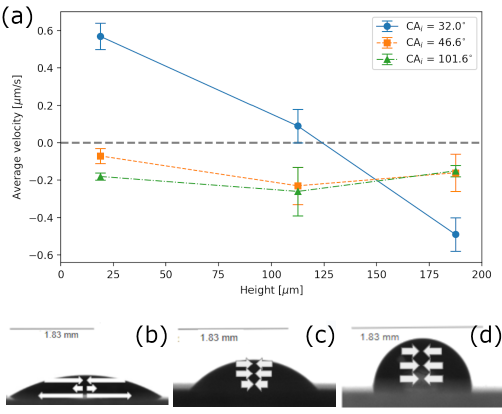}
        \caption{(a) Average velocity of particles in evaporating droplet vs height above substrate for glass, plastic and silanised glass substrates; lines are a guide to the eye. The velocity values are the averages of three repeats for each type of substrate; positive/negative velocity corresponds to outward/inward flow. (b-d) Schematic of the predominant flow of silica particles, based on panel (a), for each different substrate: (b) non-functionalised glass ($\mathrm{CA}_i = 32.0^{\circ}$), (c) plastic ($\mathrm{CA}_i = 46.6^{\circ}$) and (d) silanised glass ($\mathrm{CA}_i = 101.6^{\circ}$).}
        \label{fig:confocal3Dflow}
    \end{figure}

The fluorescent silica particles in the confocal time sequences of the evaporating droplets were analyzed at three separate stage heights above the substrate using the TrackMate plugin in ImageJ, see section \ref{sec:imganalysis}. Figure \ref{fig:confocal3Dflow}(a) shows the average in-plane velocity of particles in an evaporating droplet as a function of stage height above the substrate for glass, plastic and silanised glass substrates; by stage height we mean that we have not corrected the height above the substrate for axial distortion due to refractive index mismatch.\cite{Diel2020} For glass substrates, with low initial CA, the results are in line with literature:\cite{Rey2022} the particles move towards the edge of the droplet near the substrate, whereas they move towards the central axis of the droplet higher up in the droplet, suggesting a circulating flow (see schematic in figure \ref{fig:confocal3Dflow}(b)); the relatively strong backflow is probably due to the fixed radial position during imaging i.e.~the droplet edge moves through the field of view towards the droplet centre as stage height is increased. 

For plastic and silanised glass substrates, with intermediate to high initial CA, the average velocity suggests that particles move towards the central axis of the droplet with a relatively low velocity throughout the height of the droplet (figure \ref{fig:confocal3Dflow}(c-d)). Visually inspecting the videos for plastic and silanised glass substrates, it seems that the particles are moving inwards because the droplet edge is moving inwards. Indeed, when we subtract the radial velocity of the droplet edge from the average particle velocity (near the base of the droplet at a stage height of 19 $\mu$m above the substrate), the relative particle velocity is very close to 0 for substrates with intermediate and high initial contact angles (see ESI). This also explains why overall inward flow can still result in a coffee-ring deposit (figure \ref{fig:confocalVelocityHeight}(b) and \ref{fig:confocal3Dflow}(c)): particles' relative motion can be towards the edge, but their overall movement can be towards the centre of the droplet due to inward movement of the contact line itself. This suggests that the particle-substrate interaction is also crucial: if an initial coffee ring forms, it can induce strong pinning of the contact line, thereby suppressing further contact-line movement.

\subsection{Laser-sheet imaging}
\label{subsec:laserSheetImaging}

\begin{figure}
        \centering
            \includegraphics[width=0.99\linewidth]{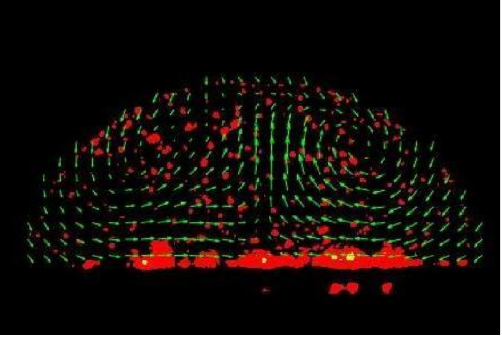}
        \caption{Laser-sheet imaging of the evaporation-induced flow ($t = 180$ s) inside of a 3 $\mu$L-water droplet containing 0.05 w/v\% PMMA spheres (1.02 $\mu$m) and 0.0005 w/v\% fluorescent polystyrene spheres (1.14 $\mu$m) placed on a silanised glass substrate. The red signal reveals the positions of the moving fluorescent tracers and the green arrows illustrate the velocity flow field. The unexpected red signal located below the substrate is an imaging artifact (glass reflections).}
        \label{fig:laserSideView}
    \end{figure}

To provide further evidence of inward flow for evaporating water droplets on substrates with relatively high contact angles, we performed laser-sheet imaging (Figure \ref{fig:laserSideView}). We find that the flow is inwards near the substrate, and recirculates near the top of the droplet. This flow persists throughout most of the evaporation process, resulting in a relatively uniform deposit at the end (see ESI). The fluorescence and confocal-microscopy measurements presented above are in agreement near the substrate, but they did not show outward flow near the top of the droplet; this is likely due to the relatively small field of view and the imaging location near the centre of the droplet. Note that some signal in Figure \ref{fig:laserSideView} is seemingly coming from within/below the substrate, but this is an imaging artefact i.e.~it is caused by reflections from the substrate--droplet interface. Overall, the laser-sheet imaging backs up a crucial aspect of the results presented above: changing the contact angle of the substrate changes the flow inside the droplet near the substrate leading to a change in the final deposit.

\subsection{Discussion}

    Combining the results presented above, a coherent explanation starts to emerge for the observed drying behaviour, its link to the internal flow within the evaporating droplets, and the corresponding changes in contact line stick-slip behaviour. For a low initial contact angle, we observe primarily a capillary outward flow near the substrate leading to the formation of an initial coffee ring, and an inward flow higher up in the droplet. The initial coffee ring induces a strong pinning of the contact line and therefore we observe a constant contact line mode (figures \ref{fig:CAvsTime} and \ref{fig:confocalVelocityHeight}).
    On the other hand, on the substrate with a high contact angle, we measure the opposite flow from the contact line towards the centre, throughout the height of the droplet. We interpret that this change in the flow profile prevents the formation of a coffee ring at the drop edge and therefore leads to a reduced pinning of contact line (figures \ref{fig:CAvsTime} and \ref{fig:confocalVelocityHeight}). This, in turn, may explain the observed constant contact angle mode and therefore the shift from a typical coffee ring deposit to a central spot-like deposit. A similar observation was made in previous research where a dot-like deposit was found for an inward flow and a typical CRE for outward flow. \cite{Malinowski2018} 
    Interestingly, for the substrate with a contact angle of $CA_i =$ 95$^{\circ}$ we observe a mixture of both behaviours. While the flow within the droplet is still primarily inwards, we observe some pinning of the contact line (figures \ref{fig:sidetop} and \ref{fig:CAvsTime}, green dotted-dashed line), but the droplet still dries into a predominantly central spot. For a contact angle of $CA_i =$ 38$^{\circ}$ and $CA_i =$ 55$^{\circ}$ the pinning is increased and the deposit is a coffee ring. This indicates that there may be a gradual transition between the two extreme drying modes. 
    We should, however, note that there were slight differences in silanising agents used to modify the wettability of substrates, for example between $CA_i$ = 38, 55 and 95$^{\circ}$ and $CA_i$ = 105$^{\circ}$, and between $CA_i$ = 102$^{\circ}$ and $CA_i$ = 105$^{\circ}$ substrates. This should be kept in mind when comparing, for example, $CA_i$ = 95$^{\circ}$ and $CA_i$ = 105$^{\circ}$ results.

\section{Conclusions}
    This article studies the role played by the initial CA in altering the evaporative flows in a sessile droplet. The fabricated substrates with systematically varied  wettability exhibit distinctive evaporation modes, identified throughout the droplet lifetime, with characteristic dried deposits found for each initial CA.
    
    Both in the experimental and DDFT modelling analysis we find very similar behaviours for the CA and base diameter versus time for intermediate initial CAs.
    The average movement of particles is more towards the centre of the drop for increased initial CA, and we have observed a concurrent shift from the expected coffee-ring-like deposit to a smaller disk-like deposit featuring a more uniform distribution of the particles; our confocal-microscopy results at different heights in evaporating droplets corroborate these results.
    Therefore, we interpret that for hydrophobic substrates the change in the flow profile prevents the formation of a coffee ring at the drop edge and that the lack of particles at the CL may, in turn, favour the slip motion of the droplet edge.
    
    An explanation for the results obtained is that for CAs larger than 90$^{\circ}$ geometrical constraints could lower the rate of evaporation at the triple phase contact line compared to the apex of the droplet.
    Therefore, particles are no longer pushed towards the edges of the drop and instead pushed towards the centre, leaving a dot-like deposit. 

    To summarise, our study suggests that the initial CA of the substrate can substantially alter the flow within drying suspension droplets and therefore alter the morphology of the resulting deposit.

\section*{Conflicts of interest}
    There are no conflicts to declare.

\section*{Acknowledgements}
    C.M.P. and J.H.J.T. thank Laura Edwards, Kathryn Knight, Marie-Capucine Pope and Martin Shaw at Croda (UK) for useful discussions.
    C.M.P. acknowledges studentship funding from the Engineering and Physical Sciences Research Council Centre for Doctoral Training in Soft Matter and Functional Interfaces (SOFI-CDT, EP/L015536/1). C.M.P.~thanks Andrew Garrie for gold coating samples for SEM.
    M.R. acknowledges the Swiss National Science Foundation Project-ID P2SKP2-194953 and the Marie Sk\l{}odowska-Curie Individual Fellowship (Grant No.101064381).
    B.D.G. thanks Andrew Archer for helpful discussions.
    S.F. and M.A.R.V acknowledge funding from the Marie Sk\l{}odowska-Curie Grant Agreement No.~955612 and project PID2020-116082GB-I00 funded by MCIN/AEI/10.13039/501100011033.
    The authors also acknowledge SOFI CDT and EPSRC for financial support for the electron microscope used in this project.
    For the purpose of open access, the author has applied a Creative Commons Attribution (CC BY) licence to any Author Accepted Manuscript version arising from this submission.


\section*{References}

\end{document}